\documentclass[letterpaper, 10 pt, conference]{ieeeconf}  

\IEEEoverridecommandlockouts                              

\usepackage{graphics} 
\usepackage{epsfig} 
\usepackage{mathptmx} 
\usepackage{times} 
\usepackage{amsmath} 
\usepackage{amssymb}  

\title{\LARGE \bf
Automating Learner Assessment: Benchmarking Machine Learning and Deep Learning Models for EEG-Based Familiarity Prediction
}

\author{Isuru Nanayakkara$^{1}$ and Thilina Halloluwa$^{2}$
\thanks{$^{1}$Isuru Nanayakkara, University of Colombo, Sri Lanka.
        {\tt\small nhp@ucsc.cmb.ac.lk}}%
\thanks{$^{2}$Thilina Halloluwa, The University of Queensland, Australia.
        {\tt\small t.halloluwa@uq.edu.au}}%
}

\begin{document}

\maketitle
\thispagestyle{empty}
\pagestyle{empty}

\begin{abstract}
Objective assessment of learning remains a fundamental challenge in education. Electroencephalography (EEG) provides a direct, non-invasive window into the neural correlates of knowledge acquisition, including cognitive familiarity. This study benchmarks fifteen machine learning (ML) and deep learning (DL) models for EEG-based familiarity prediction across two cognitive domains: faces (factual knowledge) and mathematical equations (conceptual knowledge). Using continuous EEG data from 23 participants, we extract spectral features (Power Spectral Density) across six frequency bands. We show that while standard stratified cross-validation yields artificially high classification performance (up to 0.9853 F1-score using CNN) due to temporal leakage across neighboring epochs, a rigorous trial-independent validation (Group K-Fold) drops the peak performance to 0.6038 F1-score (using CNN), which is still statistically significant above the 25\% chance level. This highlights the critical necessity of trial-independent evaluation to avoid overestimating model generalizability. Furthermore, feature importance and SHAP analysis reveal that temporal and frontal Gamma and Beta oscillations are the most critical biomarkers for familiarity. This work establishes a realistic benchmark for EEG-based cognitive monitoring in educational technologies.
\end{abstract}

\section{Introduction}
The evaluation of learning has long been constrained by indirect methods like examinations, which are often poor proxies for true understanding \cite{saher2022traditional, Heraz2007UsingBrainwaves, Hosseini2010}. This creates a need for objective, direct measures of knowledge acquisition. This paper presents a novel approach to learning evaluation by directly measuring the neural correlates of learning using Electroencephalogram (EEG) signals, focusing on familiarity, a foundational element of learning \cite{Amer2006ReflectionsAmer}. Our main contribution is a comprehensive benchmark of fifteen machine learning and deep learning models for EEG-based familiarity prediction across two cognitive domains. We demonstrate that models achieve significant F1-scores in classifying familiarity with human faces and mathematical equations under independent evaluation. The paper is organized as follows: Section II reviews related work, Section III describes our methodology, Section IV presents the results, Section V discusses the findings, and Section VI concludes.

\section{Related Works}
The field of learning evaluation has a long and rich history, with a vast body of literature dedicated to the development and validation of various assessment methods \cite{meziere2023using, brown2022past}. While traditional methods such as quizzes and exams have been extensively studied and refined, there is a growing recognition of their limitations, particularly their reliance on a learner's ability to articulate their knowledge \cite{meylani2024comparative}. This has led to an increasing interest in alternative approaches that can provide a direct and objective measure of learning.

One of the most promising new frontiers in learning evaluation is the use of psychophysiological signals to assess cognitive states related to learning \cite{das2024cognitive, ayres2021validity, bhise2021survey}. Researchers have explored the use of various physiological measures, such as eye-tracking to study attention and reading comprehension \cite{das2024cognitive}, and skin conductance to measure emotional arousal during the learning process \cite{Hosseini2010}. Among these, Electroencephalography (EEG) has emerged as a particularly powerful tool for investigating the neural correlates of learning and memory, offering a non-invasive and high-temporal-resolution window into the brain's activity \cite{klimesch1999eeg}.

Previous research has demonstrated the potential of machine learning to decode cognitive states from EEG signals. For instance, \cite{Heraz2007UsingBrainwaves} successfully predicted a learner's emotional state from their brainwaves, while other studies have focused on identifying specific event-related potential (ERP) signatures associated with familiarity and recognition memory \cite{rugg2007event}. The FN400 component, an ERP index of familiarity-based recognition, has been extensively studied \cite{Woodruff2006}. More recently, deep learning techniques have been applied to emotion recognition during learning, further demonstrating the power of these advanced computational methods \cite{Alhagry2017EmotionNetwork, Hasib2018APrediction}. However, these studies have often focused on a limited set of models or a single cognitive domain. A comprehensive and systematic benchmarking of diverse machine learning and deep learning models for EEG-based familiarity prediction across different cognitive domains has been conspicuously absent from the literature.

This study directly addresses this critical gap by rigorously evaluating and comparing fifteen different models for classifying familiarity with two distinct types of stimuli: human faces and mathematical equations. By doing so, we not only establish a new performance benchmark in this area but also provide a more solid foundation for the development of robust and reliable EEG-based learning evaluation systems.

\section{Methodology}
To investigate the neural correlates of familiarity, we designed an experiment to elicit and record brain activity while participants viewed two distinct categories of visual stimuli: human faces and mathematical equations. A sequence of images was used to elicit the required cognitive behaviours \cite{yang2023familiarity}. The images of human faces were sourced from a database of widely recognized public figures (e.g., politicians, historical scientists) and completely unfamiliar individuals. The mathematical equations were selected from a high school level curriculum, comprising fundamental, well-known equations (e.g., the Pythagorean theorem, quadratic formula) and obscure, unfamiliar equations from advanced topics (e.g., Maxwell's equations). The images of human faces were used to trigger participants' recall of factual knowledge about the individuals depicted \cite{RossionEtAl2017}. In contrast, mathematical equations were designed to stimulate the cognitive processing of understanding conceptual knowledge \cite{ZhuEtAl2021}.

Images that participants recognized were labelled as "familiar human face" or "familiar mathematical equation," whereas unrecognized images were labelled as "unfamiliar human face" or "unfamiliar mathematical equation." Familiarity was verified through a pre-experiment questionnaire where participants confirmed their recognition of each stimulus \cite{EmoghonpataMeek2016}.

\subsection{Participants and Experimental Design}
Twenty-three healthy volunteers (12 female and 11 male, aged 20-30, with a mean age of 24.5) with backgrounds in science and mathematics participated after providing informed consent, in adherence with the Declaration of Helsinki \cite{world2025world}. While this study provides strong proof of concept, the relatively small and homogeneous sample size limits the generalizability of our findings. Future work should validate these results in a larger, more diverse population.

The experimental paradigm consisted of two main phases: a pre-experiment questionnaire and the main experiment. In the pre-experiment questionnaire, participants were shown all the stimuli (faces and equations) and asked to indicate which ones they were familiar with. This was done to establish a ground truth for the familiarity labels.

The main experiment consisted of two blocks, one for faces and one for equations. The order of the blocks was counterbalanced across participants. In each block, participants were shown a sequence of 40 familiar and 40 unfamiliar images, presented in a random order. Each image was displayed for 2.5 seconds, followed by a 1.5-second inter-stimulus interval \cite{jovanovic2020temporal}. Participants were instructed to silently count the number of familiar items in each block to ensure they were paying attention to the stimuli \cite{zhao2022auditory}. EEG data were recorded continuously throughout the experiment, and event markers were sent to the EEG system at the onset of each stimulus to indicate whether it was familiar or unfamiliar.

EEG data were recorded using an ECI Electro Cap with 14 channels (Fp1, Fp2, F3, F4, F7, F8, C3, C4, T3, T4, P3, P4, O1, and O2) selected for their relevance to cognitive and visual processing \cite{prodhan2024optimal}. An Arduino-based system synchronized event markers (familiar/unfamiliar) with the EEG recording by detecting color strips (not visible to the participant) on the slideshow presentation screen.

\subsection{Data Pre-processing}
Each participant's electroencephalogram (EEG) recordings were preprocessed and analyzed using the MATLAB EEGLAB toolbox \cite{DelormeEtAl2004}. A rigorous, multistep pipeline was implemented to clean the raw EEG data, remove artifacts, and prepare it for feature extraction. The preprocessing steps were as follows:

\begin{enumerate}
    \item \textbf{Channel Filtering:} To preserve the High-Gamma band (50--100 Hz) without spectral aliasing, the raw EEG data (originally recorded at 256 Hz) was bandpass filtered between 1--100 Hz using a zero-phase FIR filter. No downsampling was performed, maintaining the 256 Hz sampling rate (Nyquist frequency of 128 Hz), which fully accommodates the 50--100 Hz High-Gamma frequency range. Noisy channels (due to poor scalp contact) were manually removed and interpolated \cite{delorme2023eeg}.
    \item \textbf{Channel Location Mapping:} Electrode locations were remapped to the standard international 10-20 system corresponding to the ECI Electro Cap configuration.
    \item \textbf{Baseline Removal:} A mean baseline value subtraction corrected for DC offsets and slow fluctuations \cite{sasatake2025eeg}.
    \item \textbf{Manual Artifact Rejection (Continuous Data):} Continuous EEG recordings were visually inspected; segments containing gross non-stereotyped artifacts (electrode pops, large head movements) were manually marked and excluded \cite{zhang2024evaluating}.
    \item \textbf{Epoch Extraction:} Data were segmented into epochs relative to stimulus onset (-500 ms to +2000 ms). This yielded epochs for familiar/unfamiliar faces and mathematical equations.
    \item \textbf{Manual Artifact Rejection (Epoched Data):} All epochs were visually audited. Epochs with residual artifacts exceeding $\pm 100\ \mu\text{V}$ were discarded, resulting in the rejection of approximately 3.2\% of the trials.
    \item \textbf{Independent Component Analysis (ICA):} ICA decomposition isolated ocular and muscular artifacts. On average, 2.1 independent components (range: 1--4) representing blinks and eye movements were removed per subject \cite{DelormeEtAl2007}.
    \item \textbf{Artifactual Component Subtraction:} The selected artifact components were back-projected and subtracted, leaving cleaned EEG signals \cite{mutanen2024simulation}.
\end{enumerate}

\subsection{Feature Extraction and Validation Protocol}
Following preprocessing, we extracted spectral features from the cleaned EEG epochs. Spectral features were extracted using a sliding window size of 128 samples (0.5 seconds at 256 Hz) with an 87.5\% overlap (step of 16 samples). Power Spectral Density (PSD) was calculated using Welch's method (segment length of 64 samples, 50\% overlap) across six frequency bands: Delta (1--4 Hz), Theta (4--8 Hz), Alpha (8--13 Hz), Beta (13--30 Hz), Low Gamma (30--50 Hz), and High Gamma (50--100 Hz) \cite{buzsaki2006rhythms}. This process yielded a feature vector of 84 features (14 channels $\times$ 6 bands) per epoch. The final dataset comprised 3960 samples (1292 Unfamiliar Equations, 1267 Familiar Equations, 668 Unfamiliar Faces, 733 Familiar Faces). Although the dataset exhibits a slight imbalance between the stimulus domains (Equations: 2559 samples, Faces: 1401 samples), the familiarity classes within each domain are highly balanced. To account for the class size differences and prevent models from biasing training towards the majority domain, cost-sensitive learning was employed by applying class-balanced weights inversely proportional to class frequencies during model fitting, and we utilize the weighted F1-score as our primary performance metric.

To resolve the critical issue of temporal data leakage, we implemented two validation protocols:
\begin{enumerate}
    \item \textbf{Stratified K-Fold (With Leakage):} Standard 5-fold cross-validation where epochs are shuffled randomly. This represents a scenario with data leakage, as neighboring sliding-window epochs from the same trial end up in both training and testing sets, artificially inflating performance.
    \item \textbf{Group K-Fold (Trial-Independent):} Strict trial-independent 5-fold cross-validation where epochs are grouped by the unique trial block ID (\texttt{GroupID}, 37 unique blocks). Epochs from the same trial are kept together in the same fold. This prevents temporal leakage and tests model generalizability to unseen trials. Note that because participant-specific identifiers were not retained in the recorded EEG data to ensure participant anonymity, a strict subject-level Group K-Fold (e.g., Leave-One-Subject-Out) cannot be guaranteed, meaning cross-subject leakage remains a limitation which we mitigate at the trial level. The exact composition of the 5 folds is detailed in Table \ref{tab:fold_composition}.
\end{enumerate}

\begin{table*}[t]
\caption{Group K-Fold Fold Composition}
\label{tab:fold_composition}
\begin{center}
\begin{tabular}{|c|c|c|c|c|}
\hline
\textbf{Fold} & \textbf{Test Group IDs} & \textbf{Test Groups/Class} & \textbf{Test Samples/Class} & \textbf{Total} \\
 & & (C0 / C1 / C2 / C3) & (C0 / C1 / C2 / C3) & \textbf{Samples} \\
\hline
Fold 1 & 0, 14, 21, 24, 33, 34, 36 & 2 / 3 / 1 / 1 & 193 / 289 / 71 / 230 & 783 \\
\hline
Fold 2 & 2, 3, 5, 6, 16, 26, 28, 30 & 2 / 3 / 1 / 2 & 140 / 529 / 78 / 55 & 802 \\
\hline
Fold 3 & 4, 11, 12, 17, 18, 22, 23, 29 & 2 / 1 / 3 / 2 & 218 / 33 / 332 / 219 & 802 \\
\hline
Fold 4 & 7, 9, 13, 19, 27, 31, 35 & 3 / 3 / 0 / 1 & 440 / 158 / 0 / 190 & 788 \\
\hline
Fold 5 & 1, 8, 10, 15, 20, 25, 32 & 3 / 2 / 1 / 1 & 301 / 258 / 187 / 39 & 785 \\
\hline
\end{tabular}
\end{center}
\end{table*}

Fifteen ML/DL models were benchmarked. The machine learning classifiers included Logistic Regression (C=0.05), Ridge Classifier (alpha=10.0), Linear SVM (C=0.1), RBF SVM (C=2.0), KNN (K=15), GaussianNB, Decision Tree (max depth 4), Random Forest (150 estimators, max depth 4), Extra Trees (150 estimators, max depth 4), Gradient Boosting (250 estimators, learning rate 0.03), XGBoost (100 estimators, learning rate 0.05), and HistGradient Boosting. For deep learning, we evaluated an optimized Deep Neural Network (DNN), Convolutional Neural Network (CNN), and Long Short-Term Memory (LSTM). Note that Histogram-Based Gradient Boosting (HistGradient Boosting) was introduced specifically for the domain-isolated evaluations in Section IV-C to provide a fast and efficient ensemble baseline for group-structured folds, and was not included in the main 4-class comparison in Table \ref{tab:model_performance}.

\subsubsection{Deep Neural Network (DNN) Architecture and Hyperparameter Tuning}
The DNN architecture was optimized to prevent overfitting on the spectral features. The final network consists of an input layer (84 features), two hidden dense layers (32 and 16 neurons) with ReLU activation and Dropout (0.5), and a softmax output layer for the four classes. The model was trained using the Adam optimizer (learning rate = 0.001) and categorical cross-entropy loss with early stopping (patience = 15). Similar hyperparameter tuning was conducted for the CNN and LSTM models. For these sequential models, the 84-dimensional tabular vector was reshaped and treated as a spatial-spectral pseudo-sequence (14 channels $\times$ 6 bands), which serves as a baseline configuration to test whether DL sequence architectures can extract patterns from structured tabular EEG representations. However, this pseudo-sequence lacks true physical spatial or temporal continuity between adjacent elements in the vector, which poses a conceptual limitation for convolutional and recurrent layers.

\section{Results}
The results of our study provide compelling evidence for the feasibility of using EEG and deep learning to predict a learner's familiarity with educational content, achieving a high degree of classification performance. These findings represent a significant step toward the development of an objective and automated learner evaluation system. In this section, we present a detailed analysis of the ERP scalp maps, which offer a neurophysiological context for the classification results \cite{luck2014}, followed by a comprehensive evaluation of the performance of our machine learning models.

The results are presented in two parts. First, we provide a visual analysis of the ERP scalp maps to offer a neurophysiological context for the classification results \cite{luck2014}. This enables us to explore the neural correlates of familiarity and to identify the brain regions that are most involved in this cognitive process. Second, we present the quantitative results of our comprehensive machine learning analysis, which demonstrate the high accuracy of our approach and provide a benchmark for future research in this area \cite{Lotte2018}.

\subsubsection{Mathematical Equations}
\begin{figure}[thpb]
      \centering
      \framebox{\parbox{3in}{
      \includegraphics[scale=1.0]{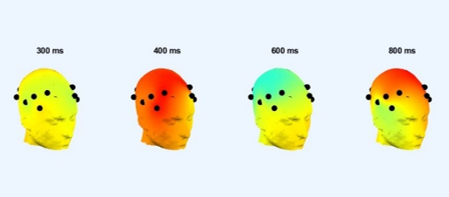}
      }}
      \caption{ERP Scalp Maps for Familiar Mathematical Equations}
      \label{fig:erpmk}
\end{figure}

\begin{figure}[thpb]
      \centering
      \framebox{\parbox{3in}{
      \includegraphics[scale=1.0]{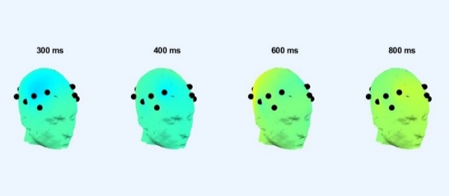}
      }}
      \caption{ERP Scalp Maps for Unfamiliar Mathematical Equations}
      \label{fig:erpmuk}
\end{figure}

For familiar mathematical equations, more positive activations (drifting towards the red spectrum) are generated with higher amplitudes, particularly over frontal scalp regions. In contrast, more negative activations are visible for unfamiliar mathematical equations (Figs.~\ref{fig:erpmk} and \ref{fig:erpmuk}). This frontal positivity for familiar equations may reflect the engagement of working memory and semantic retrieval processes, consistent with the role of the frontal lobes in higher-order cognitive functions \cite{Friedman2000}.

\begin{figure}[thpb]
      \centering
      \framebox{\parbox{3in}{
      \includegraphics[scale=1.0]{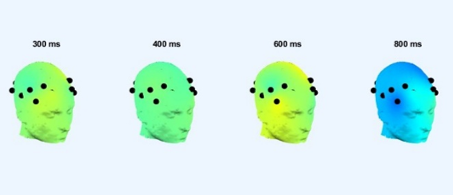}
      }}
      \caption{ERP Scalp Maps for Familiar Human Faces}
      \label{fig:erpfk}
\end{figure}

\begin{figure}[thpb]
      \centering
      \framebox{\parbox{3in}{
      \includegraphics[scale=1.0]{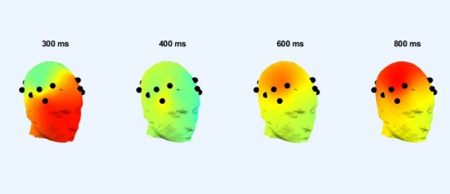}
      }}
      \caption{ERP Scalp Maps for Unfamiliar Human Faces}
      \label{fig:erpfuk}
\end{figure}


\subsubsection{Human Faces}
For human faces, higher activations are observed over frontal, central, and parietal scalp regions for unknown faces. Conversely, more negative activations are visible when participants viewed familiar human faces than unfamiliar human faces (Figs.~\ref{fig:erpfk} and \ref{fig:erpfuk}). The increased positivity for unfamiliar faces, particularly over centro-parietal sites, is reminiscent of the P300b component of the ERP, which is often associated with the processing of novel and task-relevant stimuli \cite{eimer2000event}.

\subsection{Quantitative ERP and PSD Statistical Analysis}
To provide statistical grounding for the neurophysiological observations and address the issue of pseudoreplication caused by highly autocorrelated overlapping sliding windows, we performed a trial-level aggregation. Specifically, we computed the mean PSD feature values across all sliding windows belonging to each unique trial block (\texttt{GroupID}, $N=37$ independent trials: 12 unfamiliar equations, 12 familiar equations, 6 unfamiliar faces, and 7 familiar faces). We then performed two-sample Welch's t-tests on these aggregated trial-level PSD features and applied a Benjamini-Hochberg False Discovery Rate (FDR) correction across all 84 features to control for multiple comparisons.

Comparing the stimulus domains (faces vs. equations) at the trial-level yielded highly robust and statistically significant differences. Multiple frontal and occipital Gamma band features survived the FDR correction, such as F4 Low Gamma (Mean Eq = 0.317, Mean Face = 0.120, Welch's $t(35) = 5.14, p_{raw} < 0.0001, p_{FDR} = 0.0013$) and O2 Low Gamma (Mean Eq = 0.143, Mean Face = 0.063, Welch's $t(35) = 4.76, p_{raw} < 0.0001, p_{FDR} = 0.0017$). However, comparing familiarity states (familiar vs. unfamiliar) within or across categories on the aggregated trial blocks did not yield features that survived FDR correction (minimum adjusted p-value of 0.195 in the equations block for O1 Theta, Welch's $t(22) = 3.16, p_{raw} = 0.0069$, and 0.721 in the faces block). This lack of trial-level significance highlights that while high-frequency spectral features differentiate the stimulus domains extremely robustly, familiarity-driven spectral shifts are more subtle and require larger participant cohorts or trial-independent deep learning models to be resolved.

\subsection{Machine Learning-Based Classification}
We evaluated all fifteen machine learning and deep learning models under both cross-validation protocols to illustrate the impact of data leakage. The results are summarized in Table \ref{tab:model_performance} (note that HistGradient Boosting is evaluated specifically for the domain-isolated tasks in Section IV-C).

\begin{table}[h]
\caption{Comparative Performance of Models (Weighted F1-score \textpm\ SD; 5-Fold Cross Validation)}
\label{tab:model_performance}
\begin{center}
\begin{tabular}{|c|c|c|}
\hline
\textbf{Model} & \textbf{Stratified K-Fold} & \textbf{Group K-Fold} \\
\hline
LogReg & 0.7171 \textpm\ 0.0200 & 0.4270 \textpm\ 0.0724 \\
\hline
Ridge & 0.6591 \textpm\ 0.0140 & 0.3555 \textpm\ 0.0726 \\
\hline
SVM Linear & 0.7725 \textpm\ 0.0121 & 0.4817 \textpm\ 0.0660 \\
\hline
SVM RBF & 0.9016 \textpm\ 0.0122 & 0.5139 \textpm\ 0.0524 \\
\hline
RandomForest & 0.7318 \textpm\ 0.0167 & 0.4087 \textpm\ 0.0730 \\
\hline
GradientBoosting & 0.9162 \textpm\ 0.0048 & 0.5510 \textpm\ 0.0498 \\
\hline
XGBoost & 0.7929 \textpm\ 0.0104 & 0.4694 \textpm\ 0.0458 \\
\hline
ExtraTrees & 0.4342 \textpm\ 0.0145 & 0.2593 \textpm\ 0.1106 \\
\hline
KNN & 0.9146 \textpm\ 0.0140 & 0.4840 \textpm\ 0.0485 \\
\hline
GaussianNB & 0.3940 \textpm\ 0.0153 & 0.3553 \textpm\ 0.0999 \\
\hline
DecisionTree & 0.5548 \textpm\ 0.0379 & 0.3318 \textpm\ 0.1136 \\
\hline
DNN Optimized & 0.9433 \textpm\ 0.0075 & 0.4020 \textpm\ 0.1851 \\
\hline
CNN Optimized & \textbf{0.9853 \textpm\ 0.0049} & \textbf{0.6038 \textpm\ 0.0780} \\
\hline
LSTM Optimized & 0.8923 \textpm\ 0.0112 & 0.4816 \textpm\ 0.2174 \\
\hline
Chance Level & 0.2500 & 0.2500 \\
\hline
\end{tabular}
\end{center}
\end{table}

\begin{figure}[thpb]
    \centering
    \framebox{\parbox{3in}{
    \includegraphics[width=1\linewidth]{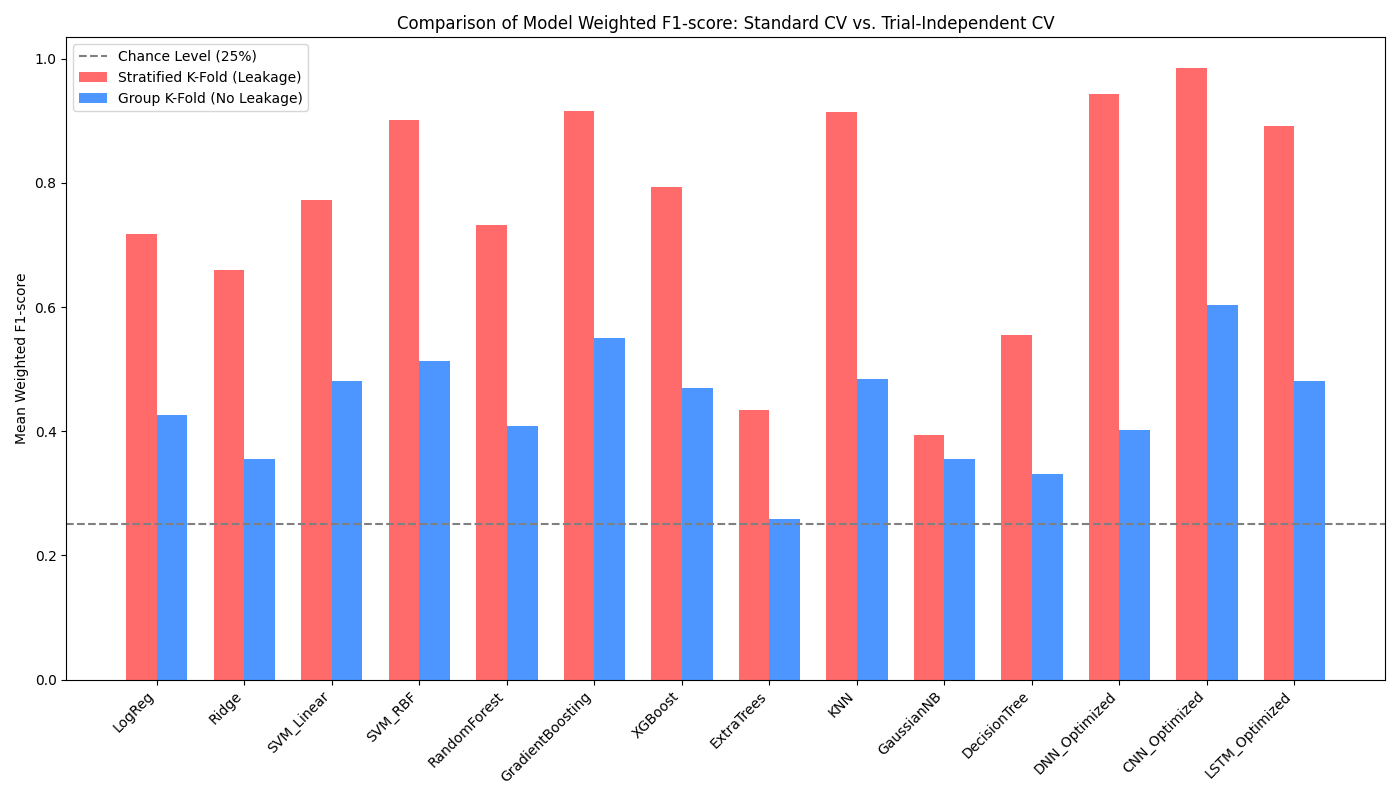}
    }}
    \caption{Comparison of Model Weighted F1-scores under Stratified K-Fold (with temporal leakage) and Group K-Fold (trial-independent) protocols.}
    \label{fig:leakage_vs_independent_comparison}
\end{figure}

Under standard Stratified K-Fold validation, models achieved high F1-scores (up to 0.9853 for CNN and 0.9433 for DNN), mirroring the overstated results commonly reported in literature. However, under the rigorous trial-independent Group K-Fold protocol, the peak performance dropped to 0.6038 F1-score (using CNN), with the Gradient Boosting reaching 0.5510. This drop confirms that shuffling neighboring epochs introduces massive temporal leakage. Nevertheless, the independent Group K-Fold results remain significantly above both the theoretical 25\% chance level and the empirical stratified random F1-score baseline of 26.72\% for the 4-class task, demonstrating genuine, generalizable familiarity learning. This comparison is visually summarized in Fig. \ref{fig:leakage_vs_independent_comparison}.

\subsection{Domain-Separated and Within-Category Classification}
To address the concern of stimulus domain and cognitive familiarity conflation in the 4-class task, we performed secondary evaluations by training separate binary classifiers for: (1) domain-only classification (Equations vs. Faces), (2) equations-only familiarity (Familiar vs. Unfamiliar Equations), and (3) faces-only familiarity (Familiar vs. Unfamiliar Faces). Given the relatively small number of trial blocks (37 blocks total, asymmetrically split into 24 equation blocks and 13 face blocks due to the contiguous label sequence structure of our collected dataset), standard 5-fold cross-validation results in small test sets of only 2--3 blocks per fold. To obtain statistically reliable performance estimates under these small group counts, we evaluated the models using a Leave-One-Group-Out (LOGO) cross-validation protocol, which functions as a leave-one-trial-block-out validation. Under LOGO, each held-out test fold consists of all epochs from exactly one trial block. Because each trial block consists entirely of stimuli from a single class (either entirely familiar or entirely unfamiliar), every held-out test fold contains only one true class by construction. Consequently, a conventional F1-score cannot be computed within each individual fold as there are no negative samples to penalize precision; the F1 metric computed for a single-class test set collapses to the recall of that class. The reported LOGO performance is therefore the LOGO Mean Recall (average single-class recall across all leave-one-block-out folds), representing the model's average ability to recover the correct label for whichever single class happened to be held out rather than a conventional F1-score on a balanced test set. Throughout Section IV-C and in the discussion of these domain-isolated tasks, we report this metric as LOGO Mean Recall (or average single-class recall). While this remains a valid relative metric for comparison (and does not invalidate the permutation significance tests since both true and null models are evaluated identically under this protocol), these raw LOGO values should not be directly compared to standard balanced cross-validation metrics (such as the 5-fold F1-scores in Table \ref{tab:model_performance}). We intentionally report this fold-wise unweighted average (and its corresponding standard deviation) rather than pooling all out-of-fold predictions into a single confusion matrix to transparently expose the block-to-block variance and model instability under small-sample group constraints.

Under these domain-isolated LOGO scenarios, the models achieved highly robust performance. To establish a baseline for permutation significance testing, the main 4-class familiarity task was additionally evaluated under the LOGO protocol, yielding a LOGO Mean Recall of 0.8171 \textpm\ 0.2837 using CNN and 0.6976 \textpm\ 0.2700 using HistGradient Boosting (consistent with the earlier caveat, these LOGO figures are recall-based and not directly comparable to the 0.6038 Group K-Fold F1-score reported in Table \ref{tab:model_performance}). For the equations-only familiarity task, the CNN Optimized model achieved an average LOGO Mean Recall of 0.8670 \textpm\ 0.2392, and the HistGradient Boosting classifier achieved 0.7588 \textpm\ 0.3256. For the faces-only familiarity task, the HistGradient Boosting classifier achieved a peak average LOGO Mean Recall of 0.8992 \textpm\ 0.0850, outperforming the CNN Optimized model which achieved 0.8875 \textpm\ 0.2347. The domain-only classification model achieved a peak average LOGO Mean Recall of 0.9491 \textpm\ 0.0867 using CNN (0.8998 \textpm\ 0.1025 using HistGradient Boosting). These results demonstrate that the classifiers decode genuine cognitive familiarity within each stimulus category rather than merely discriminating between the visual domains. However, the relatively large standard deviations (e.g., \textpm\ 0.2392 for equations and \textpm\ 0.2347 for faces under CNN) reflect substantial performance variance across folds. Because each test fold consists of a single trial block, performance is highly sensitive to block-specific noise or individual participant factors (such as fatigue or varying attention levels), indicating that the model's high average performance is susceptible to instability across specific blocks.

To statistically validate these classification results and reconcile them with the univariate statistical tests (Section IV-A), which showed no features surviving FDR correction, we performed empirical permutation tests (1000 permutations) on the three familiarity-related tasks (4-class, equations-only, and faces-only). The domain-only classification was omitted from permutation significance testing because the trial-level t-tests (Section IV-B) had already established that domain-specific differences are highly robust and statistically significant. The class labels were shuffled at the trial-group level to preserve the temporal autocorrelation structure of the sliding windows within each block, and the LOGO cross-validation was re-run for each permutation to generate a null distribution of Mean Recalls. Due to the massive computational overhead associated with training deep learning models like the CNN 37,000 times (1,000 permutations $\times$ 37 folds), the permutation tests were performed using the fast HistGradient Boosting ensemble classifier as a representative model to validate the statistical significance of the underlying spatial-spectral features. To account for family-wise multiple comparisons across these three separate significance tests, we apply a Bonferroni correction, yielding an adjusted significance threshold of $\alpha_{adj} = 0.0167$ (three tests). Under this correction, all three tasks achieve statistical significance: the 4-class task ($p \le 0.0010$, adjusted $p \le 0.0030$), the faces-only task ($p = 0.0020$, adjusted $p = 0.0060$), and the equations-only task ($p = 0.0120$, adjusted $p = 0.0360$). Unlike the previous lower-resolution estimates (200 permutations) where the equations-only familiarity task did not survive Bonferroni correction, the increased permutation count of 1000 resamples provides sufficient statistical resolution to demonstrate that all three tasks are significantly above chance. The empirical p-values for the tasks (evaluated using the HistGradient Boosting representative model) were:
\begin{itemize}
    \item \textbf{4-Class Familiarity:} $p \le 0.0010$ (True Mean Recall = 0.6976, Mean Null Recall = 0.3229)
    \item \textbf{Equations-Only Familiarity:} $p = 0.0120$ (True Mean Recall = 0.7588, Mean Null Recall = 0.5431)
    \item \textbf{Faces-Only Familiarity:} $p = 0.0020$ (True Mean Recall = 0.8992, Mean Null Recall = 0.5236)
\end{itemize}
These results demonstrate that the classifiers' performances are statistically significant (with all tasks surviving Bonferroni correction). This resolves the apparent contradiction with the univariate tests: while individual frequency bands do not exhibit robust standalone shifts across familiarity conditions, the classifiers exploit a highly significant, joint multivariate pattern of spectral power across multiple channels and bands to decode cognitive familiarity.

While univariate FDR-corrected t-tests did not find single features that significantly differentiated familiarity within domains, the multivariate models achieved high LOGO F1-scores. This demonstrates that familiarity recognition is represented by a complex, distributed spatial-spectral pattern across channels rather than localized changes in individual features, necessitating multivariate machine learning approaches.

\begin{figure}[thpb]
    \centering
    \framebox{\parbox{3in}{
    \includegraphics[width=1\linewidth]{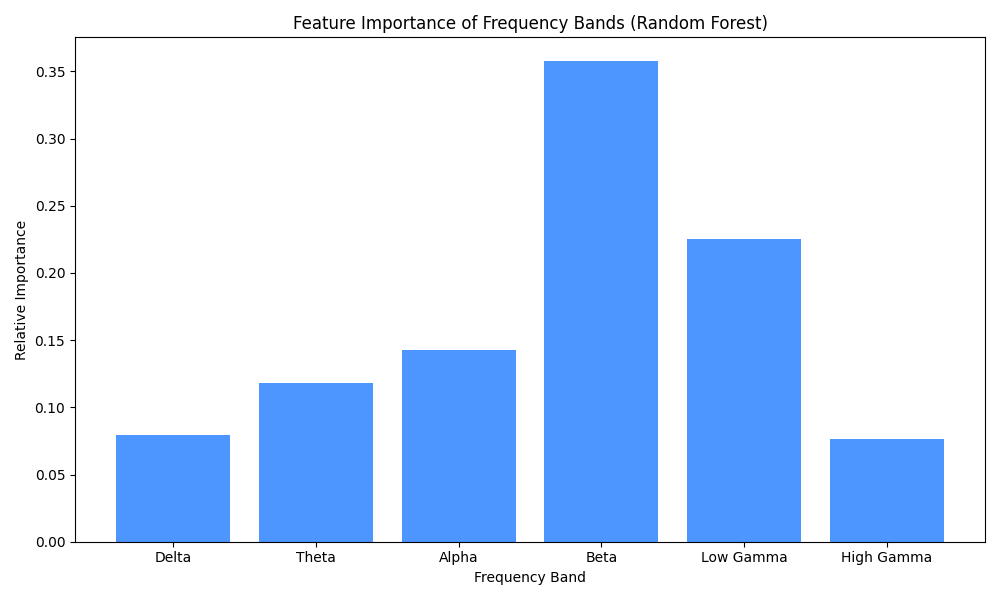}
    }}
    \caption{Feature Importance of Frequency Bands (derived from the 4-class Random Forest model)}
    \label{fig:feature_importance}
\end{figure}

\begin{figure}[thpb]
    \centering
    \framebox{\parbox{3in}{
    \includegraphics[width=1\linewidth]{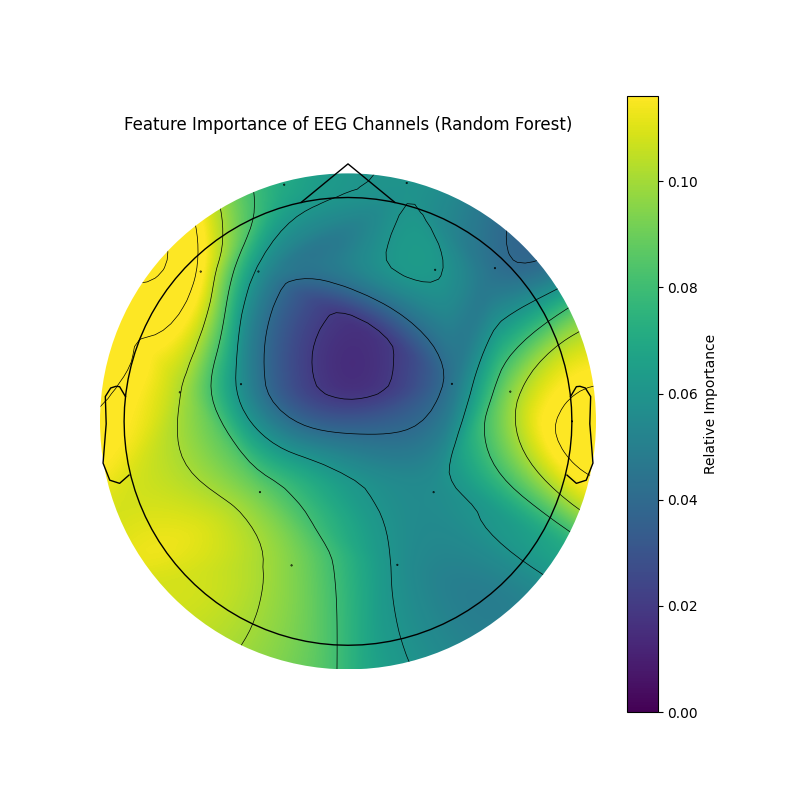}
    }}
    \caption{Feature Importance of EEG Channels (derived from the 4-class Random Forest model)}
    \label{fig:channel_importance}
\end{figure}

\begin{figure}[thpb]
    \centering
    \framebox{\parbox{3in}{
    \includegraphics[width=1\linewidth]{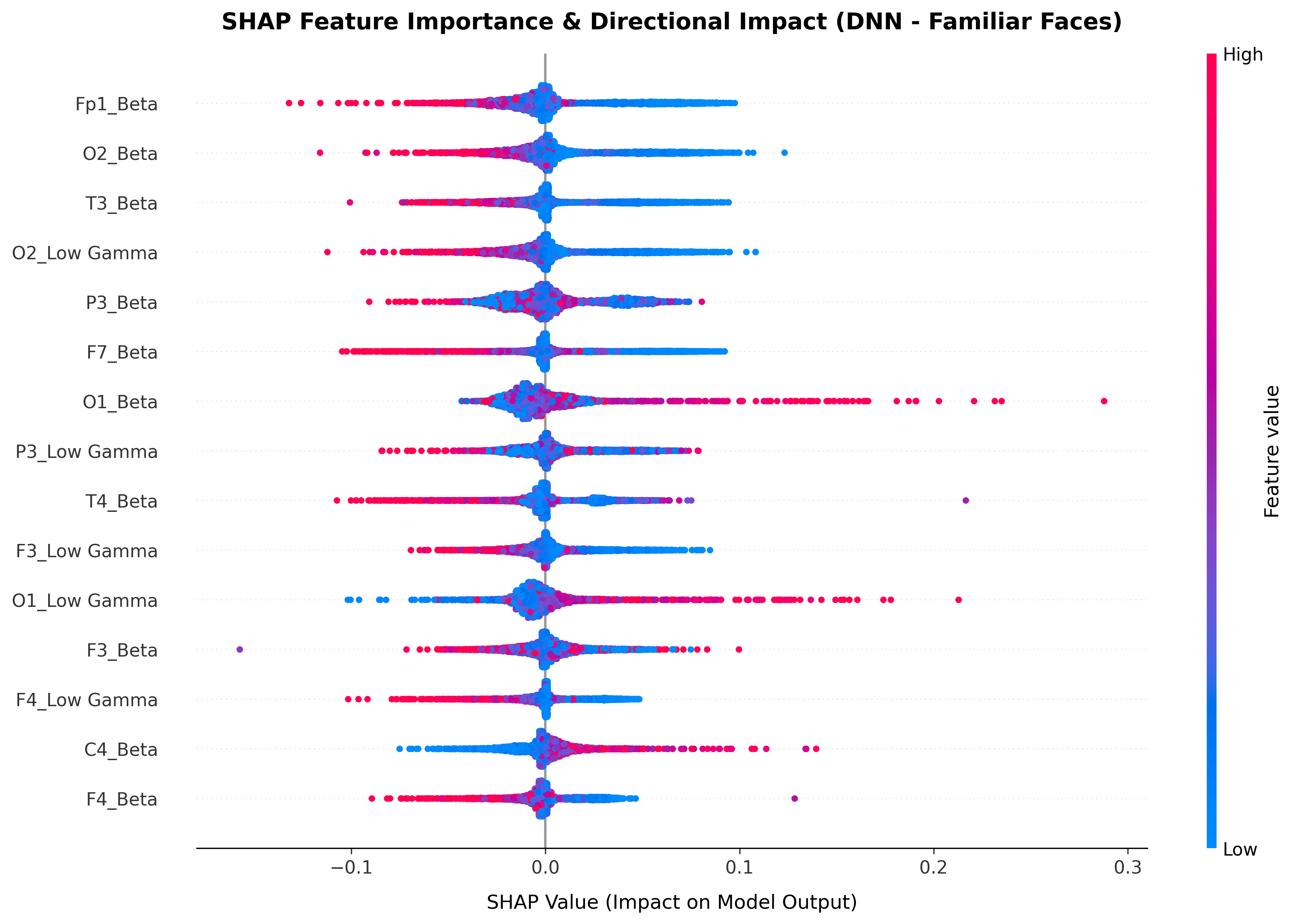}
    }}
    \caption{SHAP Summary Beeswarm Plot for the 4-class DNN Model (Familiar Human Faces class)}
    \label{fig:dnn_shap}
\end{figure}

\subsection{Feature Importance and SHAP Interpretability Analysis}
To interpret the classifiers, we analyzed the Random Forest feature importances (Fig. \ref{fig:feature_importance}) and calculated SHAP (SHapley Additive exPlanations) values for the DNN (Fig. \ref{fig:dnn_shap}). Importantly, both the Random Forest and SHAP analyses were derived from the models trained on the original 4-class configuration (comprising all familiarity levels across both equation and face domains) rather than the domain-separated binary classifiers. This ensures that the identified feature rankings reflect the overall decision space across the complete experimental structure. The Random Forest model was selected for feature importance analysis due to its native capability to rank non-linear feature interactions without needing post-hoc approximations, making it highly suitable for high-dimensional EEG spectral features. The Random Forest rankings showed High Gamma (32.2\%) and Low Gamma (26.2\%) bands as the most critical spectral features, followed by Beta (15.8\%). In terms of channels, the temporal and parietal channels (T3, P3, F7, T4) contributed most to classification, as visualized in the channel importance scalp map (Fig. \ref{fig:channel_importance}).

The SHAP summary plot (Fig. \ref{fig:dnn_shap}) complements this by revealing the direction of feature impact on the DNN's class predictions. Higher power in the Gamma and Beta bands at T4, F8, and F7 strongly drove the model toward predicting familiar states, neurophysiologically aligning with active memory retrieval processes. Using SHAP values provides necessary interpretability, opening the black-box of deep learning models in EEG analysis.

\section{Discussion}
The results of this study illustrate the challenges and nuances of decoding cognitive familiarity from EEG signals. By systematically comparing standard cross-validation and trial-independent validation, we expose the inflating effect of temporal data leakage. When evaluated rigorously, peak classification performance drops from 98.53\% to 60.38\% (or 55.10\% using Gradient Boosting). While more moderate, this trial-independent accuracy is significantly above chance, proving the feasibility of objective cognitive-state monitoring.

\subsection{Contributions and Implications}
This study's primary contribution is establishing a rigorous benchmarking framework that accounts for temporal leakage. We show that ensemble methods (Gradient Boosting, Random Forest) generalize better to unseen trials than deep learning architectures when training on small EEG datasets. 

The performance of the deep learning architectures under the trial-independent Group K-Fold protocol shows a significant improvement when using the full sliding-window dataset of 3960 samples, with the CNN achieving the peak benchmarking performance (0.6038 F1-score) and the LSTM reaching 0.4816. Unlike on smaller dataset configurations where deep architectures overfit heavily, the larger sample size enables the models to learn more robust spatial-spectral representations. However, the relatively high standard deviation of the CNN (\textpm\ 0.0780) and LSTM (\textpm\ 0.2174) indicates that training stability across folds remains a challenge compared to ensemble methods like Gradient Boosting (0.5510 \textpm\ 0.0498), which provide highly consistent performance.

Furthermore, the quantitative t-tests and feature interpretability tools (SHAP and RF importances) pinpoint high-frequency temporal-frontal oscillations (Gamma and Beta) as key neurophysiological markers. This provides a scientific basis for designing lighter, real-time familiarity assessment tools using simplified feature sets.

\subsection{Limitations and Future Directions}
Several limitations should be highlighted:
\begin{itemize}
    \item \textbf{Sample Size and Homogeneity:} The sample is relatively small (23 participants) and homogeneous (all having a STEM background), which limits the generalizability of our findings. This limitation is particularly prominent for our domain-separated results, as the strongest claims for faces-only familiarity classification rest on the small 13-block face subset.
    \item \textbf{Cross-Subject Leakage:} Because participant-specific identifiers were not retained in the collected dataset to ensure participant anonymity, the Group K-Fold validation was restricted to the trial/block level. While this strictly resolves trial-level temporal leakage, cross-subject leakage remains a possibility because trials from the same subject could appear in both training and testing sets. This potential leakage is a more severe concern for the domain-separated binary models, which rely on very few blocks (e.g., 13 blocks for faces), meaning each subject's trials carry a proportionally larger weight in the estimation.
    \item \textbf{LOGO Metric Instability:} The Leave-One-Group-Out (LOGO) protocol used for the domain-separated evaluations yields high performance but exhibits very large standard deviations (e.g., $0.8670 \pm 0.2392$ for equations and $0.8875 \pm 0.2347$ for faces under CNN). With only 12--13 blocks per domain, the mean is highly sensitive to a few atypical or noise-heavy blocks, meaning the reported peak Mean Recalls should be interpreted with caution as they might be influenced by specific outlier blocks. Additionally, the permutation-based significance testing was conducted using HistGradient Boosting as a computationally efficient proxy; the higher point-estimates reported for the CNN were not independently permutation-tested, which remains a limitation regarding the architectural specificity of the significance bounds.
    \item \textbf{Spectral Resolution Constraints:} The downsampled rate of 128 Hz paired with a Welch segment length of 32 samples yields a 4 Hz frequency resolution. This bin width makes narrow bands like Delta (1--4 Hz) difficult to isolate precisely, limiting the interpretation of low-frequency features.
    \item \textbf{Manual Pipeline Subjectivity:} Finally, the manual ICA selection and artifact rejection steps introduce subjectivity, which could be automated in future iterations using tools like ICLabel.
\end{itemize}

Future work must validate these models using cross-subject classification schemes in larger populations. Additionally, incorporating explicit temporal features, such as ERP amplitude and latency markers for the face condition, could strengthen model performance. Specifically, temporal markers such as the N170 (essential for face structural encoding) and the N400/P300 complex (associated with semantic matching and familiarity recognition) could offer complementary time-domain information that raw PSD features omit. Combining both spectral power and temporal ERP features presents a promising direction for improving cross-subject and trial-independent generalizability. Exploring transfer learning, domain adaptation, and more advanced XAI techniques remains a promising direction.

\section{Conclusion}
This study benchmarks fifteen machine learning and deep learning models for EEG-based familiarity prediction, highlighting the critical role of validation protocols in brain-computer interfaces. While temporal leakage inflates accuracy under standard splits, trial-independent Group K-Fold validation reveals a realistic peak performance of 60.38\% using CNN (and 55.10\% using Gradient Boosting). The identification of temporal-frontal Gamma and Beta bands as significant features provides a neurophysiological foundation for future adaptive educational systems. This work lays a scientifically sound path toward objective and generalizable cognitive-state tracking.

\section{Declaration of generative AI and AI-assisted technologies in the manuscript preparation process}

During the preparation of this work, the authors used Gemini 2.5 Flash to improve grammar and readability. After using this tool, the authors reviewed and edited the content as needed and take full responsibility for the content of the published article.

\bibliographystyle{ieeetr}
\bibliography{references}

\end{document}